# Cooper – like paring and energy gap induced by ion electronic polarizability.

Yizhak Yacoby and Yakov Girshberg

Racah Institute of physics, Hebrew University, Jerusalem, Israel 91904

We explore the possibility that the ionic electron polarizabilities of the oxygen ions in the cuprates and the bismutates and the polarizabilities of As and Se ions in the iron pnictides contribute to charge carrier pairing leading to high Tc superconductivity. Using the fact that the ionic polarization response to a change in the electric field is practically instantaneous we find, that the inter carrier electrostatic potential is attractive in a limited distance range. This potential is used to calculate quantum mechanically the cooper-like pairing energy and wavefunction and the gap energy showing they are consistent with pairing and gap energies of high Tc superconductors. Qualitative considerations suggest that this model may explain a number of important features of high Tc superconductors.

1. Introduction.

The discovery of superconductivity in $LaSrCuO_4$ by Bednorz and Müller[1] was very surprising to most physicists excluding perhaps the discoverers. In 1986 after decades of intensive research the highest superconducting phase transition found was 23.2K[2] in spite of the fact that the mechanism responsible for superconductivity had been discovered[3]. This largely unexpected discovery and the new class of materials involved conceptually revolutionized the field. Since this discovery two additional material classes have been found to exhibit high Tc, the Bismutates[4-6] and the iron pnictides[7,8]. These classes differ in some of their properties. For example the cuprates undergo at low doping level an antiferromagnetic transition[9]. The Bismutates and pnictides do not exhibit such a transition. This raises a very interesting question: is the underlying superconducting mechanism common to all three classes.

It is generally agreed that the phase transition to a superconducting state is due to the presence of a correlated system of charge carrier pairs (Cooper or Cooper-like pairs) [10]. It is agreed, that in all cases, even in the most exotic, their occurrence is the result of the carriers interacting with bosonic excitations of the lattice or the electronic subsystem or both [11]. These excitations can be phonons in the BCS theory [12,13], or excitons, plasmons, and antiferromagnetic spin fluctuations in BCS-like models[11,14]. In the case of strong coupling the carriers transform to polarons (lattices or spin polarons) and the polaron-polaron interaction causes the formation of bipolarons leading to superconductivity (Hubbard-like models)[15]. Very many models have been proposed but to date there is no generally accepted theory of high Tc superconductivity.

In this paper we consider a model with a different paring mechanism. We assume that the lattice is frozen with fixed ions and free carriers, interacting exclusively via ion electronic polarizability. The ion electronic polarizability is defined as the ratio between the ion dipole moment divided by the electric field at the ion center induced by the surrounding charges excluding the nuclear and electronic charges of the ion itself. The oxygen in the cuprates and bismutates and the As and Se in the iron pnictides are known to have large electron polarizabilities [16, 17]. In the framework of this model we have constructed the potential and wavefunction of two carriers interacting with each other and with the surrounding polarizable ions. We have numerically solved the Cooper problem in the presence of N other carriers showing that the lowest pair energy is lower than twice the Fermi energy. We then used the BCS wavefunction and by minimizing the total energy, calculated the gap energy $\Delta$ as a function of doping and compared it with experiment.

The effects of ionic polarizability have been discussed in several papers, but from a different point of view. In reference [18] the authors discussed the effect of polarizability on the electron – breathing mode interaction, which involves movement of the apical oxygen in a double-potential-well. In reference [19,20] the authors calculated the renormalized electron – phonon coupling constant, due to the ion electronic polarizability of the medium. In reference [21] the authors discuss the renormalization of the parameters in multiband Hubbard model caused by medium polarizability. In essence these works discuss the effect of polarizability on the electron phonon interaction. In contrast in our model the ions are frozen namely we neglect the electron phonon interaction altogether.

Recently Sawatzky et. al.[22] and Berciu et al.[23] suggested that electronic polarizability of the As or Se ions in the iron pnictides can give rise to the formation of polaron like states and these polarons attract each other creating bi-polarons located at nearest Fe positions. The authors suggest that these Bosons exist above Tc and at low temperatures undergo Bose-Einstein condensation leading to super-conductivity. The authors suggest that this mechanism does not work in the cuprates but may be the pairing mechanism in the iron pnictides. In this calculation the authors have omitted the direct Coulomb repulsion between the two charge carriers and point out that the repulsion energy involved is large and may split the bi-polarons. In our model the carrier-carrier interaction via ion polarizability creates Cooper pairs. These pairs have positive energy namely they are not stable free bozons above Tc but their energy (including repulsion) is smaller than the two particle energy at the Fermi level.

The rest of the paper consists of the following sections. In section 2, we discuss the carrier pair potential energy and the wavefunction of the Cooper like pairs. In section 3 we determine the BCS like wavefunction and determine the energy gap as a function of doping. Some of the properties of this model are qualitatively discussed in section 4 while in section 5 we summarize the results and point out further work that needs to be done.

## 2. Pairing of two charge carriers at the Fermi level induced by ionic polarizability

To illustrate the idea we treat a simple perovskite system as shown in Fig. 1. We have chosen the following oxide systems typical parameters. The unit cell size equals 0.4nm; the effective mass equals $3m_0$ and the Fermi energy is about 0.125eV. The time it takes a carrier at the Fermi level to move from one cell to the next is about $10^{-15}$ sec. The $O^{2-}$ polarizability is typically 5Å$^3$ [16]. Using this value one can estimate the response time of the $O^{2-}$ ion to an abrupt change in the electric field at its center to be about $10^{-16}$ sec. We can therefore assume that the polarizability response is instantaneous. A carrier at point $\vec{r}_1$ induces strong polarization in the nearby ions which in turn induce an electrical potential $\varphi_p(\vec{r}_1,\vec{r}_2)$ at point $\vec{r}_2$. This potential includes the effect of screening by the other free carriers. If $|\vec{r}_2 - \vec{r}_1|$ is much larger than the inter-ionic distances the potential at $\vec{r}_2$ is always repulsive or zero. However as shown below at distances $|\vec{r}_2 - \vec{r}_1|$ comparable to inter-ionic distances the potential can be attractive within limited regions in space.

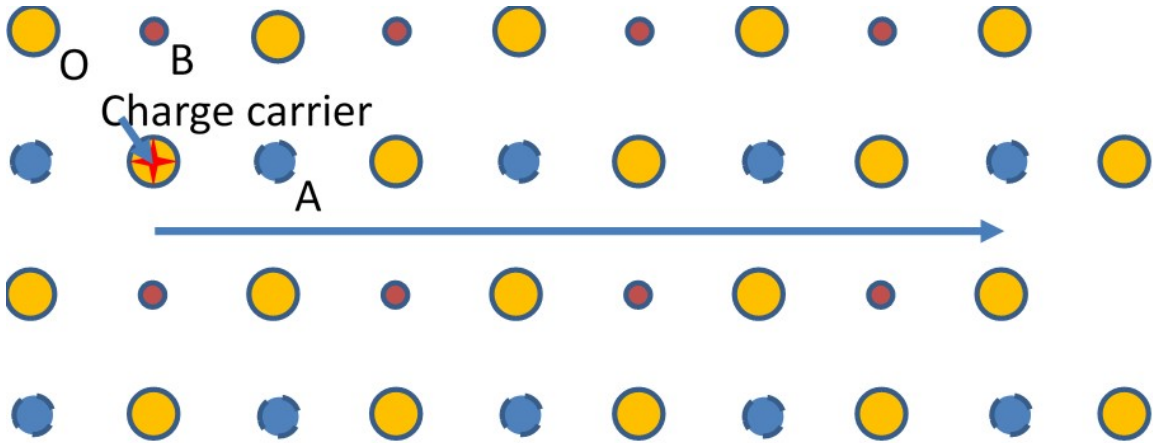

Figure 1. Perovskite like structure. The potential shown in Fig. 3 is calculated along the arrow.

Notice that this potential is not periodic in either $\vec{r}_1$ or $\vec{r}_2$. However the potential induced by the polarized ions at $\vec{r}_2 = \vec{r}_1$ is periodic in $\vec{r}_1$.

Thus, the one carrier Hamiltonian can be expressed in the following form:

$$H_{(1)} = \frac{p^2}{2m} + e\varphi_0(\vec{r}_1) + e\varphi_p(\vec{r}_1, \vec{r}_1) \tag{1}$$

Here $\varphi_0(\vec{r}_1)$ is the mean field potential at point $\vec{r}_1$ excluding the contribution of the polarization of the nearby ions. The wavefunctions that solve the corresponding one carrier Schrödinger equation are Bloch functions with wave vectors as good quantum numbers and in contrast to the interaction with phonons they do not form localized polarons. The reason is that due to the practically instantaneous polarizbility response the charge carrier 'drags' the potential well with it.

Let us now consider the Hamiltonian in the presence of 2 carriers. Assuming the polarizability coefficient is constant independent of the electric field.

$$H_{(2)} = \frac{p_1^2}{2m} + \frac{p_2^2}{2m} + e\varphi_0(\vec{r}_1) + e\varphi_0(\vec{r}_2) + e\varphi_p(\vec{r}_1, \vec{r}_1) + e\varphi_p(\vec{r}_2, \vec{r}_2) + e\varphi_p(\vec{r}_1, \vec{r}_2) + e\varphi_R(|\vec{r}_1 - \vec{r}_2|) \tag{2}$$

Here $\varphi_R(|\vec{r}_1 - \vec{r}_2|)$ is the repulsive potential between the two carriers taking screening into account.

We define $\vec{S} = (\vec{r}_2 - \vec{r}_1)/2$ and $\vec{R} = (\vec{r}_2 + \vec{r}_1)/2$

So $\varphi(\vec{R}, \vec{S}) = \varphi_p(\vec{r}_1, \vec{r}_2) + \varphi_R(|\vec{r}_1 - \vec{r}_2|)$ is periodic in $\vec{R}$, is not periodic in $\vec{S}$ and vanishes at large $S$. If $\varphi(\vec{R}, \vec{S})$ is positive for all values of $\vec{R}$ and $\vec{S}$ it will not contribute to pairing but as shown below $\varphi(\vec{R}, \vec{S})$ can be negative in limited regions of $\vec{S}$, giving rise to pairing.

The potential $\varphi(\vec{R}, \vec{S})$ can be calculated by solving the equations presented in appendix 1. These equations take into account screening in the gellium approximation and calculate the potential induced by the ionic polarizability self consistently. The ratio between the bare and screened electric field of one charge carrier is shown in Fig. 2

As an example we have calculated the electrical potential under the conditions stated above and the oxygen polarizability equals $6.56 \times 10^{-40}$ Cm$^2$/V $= 5.904$Å$^3$. The potential as a function of position along the arrow shown in Fig. 1 is shown in Fig. 3.

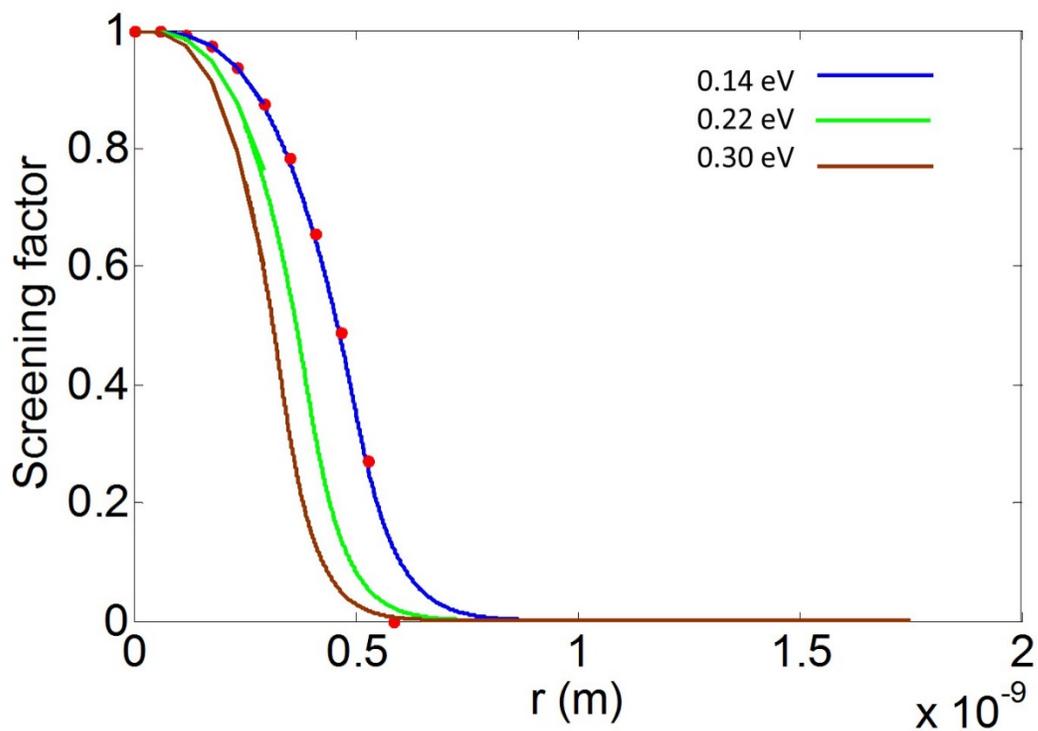

Figure 2. Screening factor as a function of distance r(m) for 3 Fermi energies 0.14, 0.22, 0.30 eV that correspond to 0.1, 0.21, 0.33 carriers per unit cell doping levels.

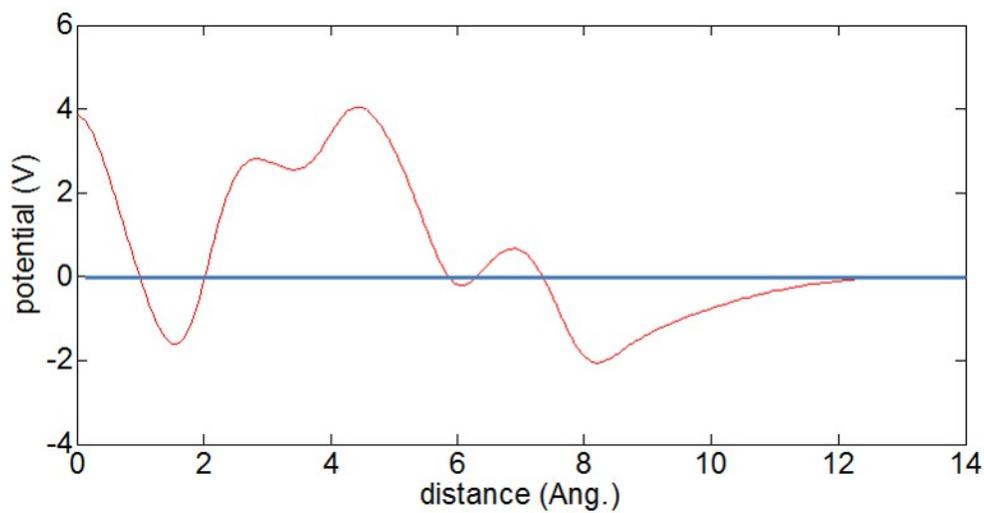

Figure 3. The electrical potential induced by the charge carrier along the arrow shown in Fig. 1.

Notice that the potential is negative between 0.1-0.2nm and 0.73-1.2nm. The details of the potential function depend on the various parameters but the presence of negative potential in regions of the order of 0.2 nm persist provided the polarizability is large enough. The fact that the potential is negative and in fact rather large negative does not mean that the two carriers form a bound state. The reason is that to form a bound state the distance between them must be confined to a very narrow region. Due to uncertainty this requires a very large kinetic energy. In fact to confine the pair distance uncertainty to 0.1nm requires an energy of the order of 1eV. Namely the negative potential and kinetic energies are of the same order of magnitude.

The one carrier wave functions are the solutions of:

$$H_{(1)} \Psi_{\vec{K}}(\vec{r}) = \varepsilon_{\vec{K}} \Psi_{\vec{K}}(\vec{r}); \tag{3}$$

Where $\Psi_{\vec{K}}(\vec{r})$ is the Bloch wavefunction.

$$\Psi_{\vec{K}}(\vec{r}) = \phi_{\vec{K}}(\vec{r}) \exp(i\vec{K} \bullet \vec{r}) \tag{4}$$

We express the solution of the 2 carrier Schrödinger equations in terms of the Bloch functions. Since the states with $K, L < K_F$ are all occupied we construct the pair wavefunction with $K, L > K_F$.

$$\Psi_{(2)}(\vec{R}, \vec{S}) = \sum_{K,L} U_{K,L} \phi_K(\vec{R} - \vec{S}) \phi_L(\vec{R} + \vec{S}) \exp(i(\vec{K} + \vec{L}) \bullet \vec{R}) \exp(i(\vec{L} - \vec{K}) \bullet \vec{S}) \tag{5}$$

Where, $U_{K,L}$ are the expansion coefficients.

Notice that since $H_{(2)}$ is periodic in $\vec{R}$, so $\vec{K} + \vec{L} = const$. As usual we expect that the pair with $\vec{K} + \vec{L} = 0$ will have the lowest energy.

We can now express the Hamiltonian in second quantization form:

$$H = \sum_{\vec{K}} (\xi_{\vec{K}} c^+_{K\uparrow} c_{K\uparrow} + \xi_{\vec{K}} c^+_{-K\downarrow} c_{-K\downarrow}) + \sum_{\vec{K},\vec{K}'} \Phi(\vec{K}, \vec{K}') c^+_{-K'\downarrow} c^+_{K'\uparrow} c_{K\uparrow} c_{-K\downarrow} \tag{6}$$

Where $\xi_{\vec{K}} = \varepsilon_{\vec{K}} - E_F$ and $c^+_{K\uparrow}, c^+_{-K\downarrow}$ create a carrier with the corresponding Bloch wavefunction and spin up or down, respectively.

And the wave function consisting of N carriers plus 2

$$\left|\Psi_{(2)}\right\rangle = \prod_{K<=K_F} c^+_{-K\downarrow} c^+_{K\uparrow} \sum_{K>K_F} U(\vec{K}) c^+_{-K\downarrow} c^+_{K\uparrow} \left|0\right\rangle \tag{7}$$

And

$$\Phi(\vec{K},\vec{K}') = \iint d^3R\, d^3S\, \phi_{K'}^*(\vec{R}+\vec{S})\, \phi_{-K'}^*(\vec{R}-\vec{S}_1)\, e\varphi(\vec{R},\vec{S})\, \phi_{-K}(\vec{R}-\vec{S})\, \phi_K(\vec{R}+\vec{S})\exp[i2(\vec{K}-\vec{K}')\bullet \vec{S}]$$

(8)

Notice that H is similar to the BCS Hamiltonian[12] except that $\Phi$ cannot be approximated by a negative constant and is given by equation 8.

As usual the Schrödinger equation

$$H|\Psi_{(2)}\rangle = (\Xi_{(0)} + \Xi_{(2)})|\Psi_{(2)}\rangle \qquad (9)$$

where $\Xi_{(0)} = \sum_{K<K_F} 2\xi_{\vec{K}}$

can be solved by solving the matrix equation :

$$\sum_{K>K_F}[\Phi(\vec{K},\vec{K}') + (2\xi_K - \Xi_{(2)})\delta_{\vec{K}\vec{K}'}]U(\vec{K}) = 0 \qquad (10)$$

Evaluating $\Phi$ properly is very complicated so to illustrate the idea we have used plane waves instead of the Bloch functions and evaluated numerically the following function:

$$\Phi(\vec{K},\vec{K}') = \iint d^3R\, d^3S\; e\varphi(\vec{R},\vec{S})\exp[i2(\vec{K}-\vec{K}')\bullet \vec{S}] \qquad (11)$$

This matrix has 3294x3294 elements.

Substituting this function in Eq. 10 we have solved the equation numerically. This equation has a large number of solutions, one of which has the lowest energy. If its energy is less than the Fermi energy namely $\Xi_{(2)} < 0$ it constitutes a pair in the sense of a Cooper pair. The absolute value of the pairing energy as a function of polarizability for a doping level that corresponds to 0.14eV Fermi energy is shown in Fig. 4. Below 4.9Å$^3$ the pairing energy is about zero. Above 6.3 Å$^3$ the electric field and potential do not converge self-consistently. The pairing energies corresponding to the polarizability levels 5-6.3 Å$^3$ are comparable to the high Tc superconductor gap and pseudo gap energies.

We have also calculated the corresponding pair wavefunctions. As an example we show in Fig. 5 the absolute value of the pair wavefunction in a system with Fermi energy $E_F = 0.14 eV$ and polarizability $\chi = 5.76 Å^3$. It shows that the distance between the two carriers is about 1nm namely much smaller than the distance in classical Cooper pairs.

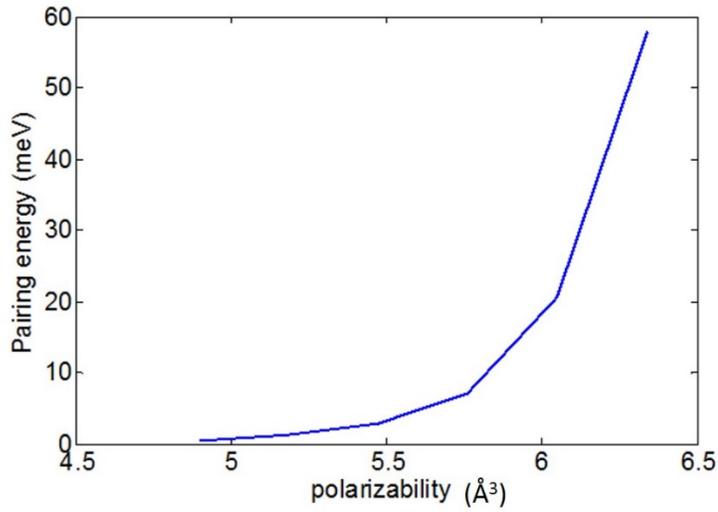

Figure 4. Pairing energy as a function of polarizability

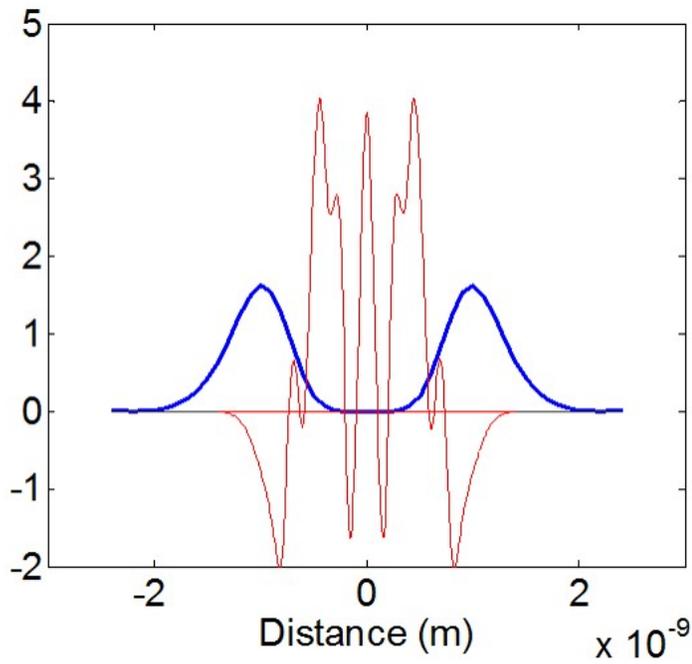

Figure 5. The absolute value squared of the wave function along the arrow (blue); The potential that the second carrier sees along the arrow (red).

3. **The energy gap**

Let us consider now the BCS wave-function[12]

$$|\Psi\rangle = \prod_{\vec{K}} (u_{\vec{K}} + v_{\vec{K}} c^+_{-K\downarrow} c^+_{K\uparrow}) \quad u_{\vec{K}}^2 + v_{\vec{K}}^2 = 1 \tag{12}$$

Here $u_{\vec{k}}, v_{\vec{k}}$ are numerical coefficients.

The Hamiltonian is given by

$$H = \sum_{\vec{K}} (\xi_{\vec{K}} c^+_{K\uparrow} c_{K\uparrow} + \xi_{\vec{K}} c^+_{-K\downarrow} c_{-K\downarrow}) + \sum_{\vec{K},\vec{K}'} \Phi(\vec{K},\vec{K}') c^+_{-K'\downarrow} c^+_{K'\uparrow} c_{K\uparrow} c_{-K\downarrow} \tag{13}$$

$\Phi(\vec{K},\vec{K}')$ as calculated in Eq. 8 and 11 is not isotropic and may contribute to d wave behavior. However to simplify the calculation discussed below we have symmetrized it so the results are s like.

The total energy can be expressed in terms of $u_{\vec{k}}, v_{\vec{k}}$

$$\Xi = \sum_{\vec{K}} 2\xi_K v_K^2 + \sum_{K,K'} \Phi(K,K') u_K v_K u_{K'} v_{K'} \tag{14}$$

$u_{\vec{k}}, v_{\vec{k}}$ can be expressed in terms of the gap $\Delta_K$ and the excitation energy $E_K = (\xi_K^2 + \Delta_K^2)^{1/2}$

$$\Xi = \sum_{\vec{K}} \xi_{\vec{K}} (1 - \frac{\xi_{\vec{K}}}{E_{\vec{K}}}) + \frac{1}{4} \sum_{\vec{K},\vec{K}'} \Phi(\vec{K},\vec{K}') \frac{\Delta_{\vec{K}} \Delta_{\vec{K}'}}{E_{\vec{K}} E_{\vec{K}'}} \tag{15}$$

The solution to the Schrödinger equation with the lowest energy is obtained by adjusting $\Delta_K$ to minimize $\Xi$. $E_K$ and $\xi_K$ are very large compared to $\Delta_K$ except close to $K_F$ so we can limit the sum to a region close to the Fermi surface and assume that $\Delta_K$ is approximately constant in the direction perpendicular to the Fermi surface but in general it depends on the direction. For simplicity we assume in this work that the system is isotropic. So $\Phi$ depends on $|\vec{K}_F|$ and on the angle between $\vec{K}_F$ and $\vec{K}_F'$ namely $\theta$ (see Fig. 6). So the total energy is approximately:

$$\Xi = 4\pi D \int_{K_F - \Delta K}^{K_F + \Delta K} dK K_F^2 \xi_K (1 - \frac{\xi_K}{E_K}) +$$

$$2\pi^2 D^2 \int_{K_F - \Delta K}^{K_F + \Delta K} dK K_F^2 \int_{K_F - \Delta K}^{K_F + \Delta K} dK' K_F^2 \frac{\Delta_{K_F}^2}{E_K E_{K'}} \int_0^\pi \Phi(\vec{K}_F, \vec{K}_F') \sin(\theta) d\theta$$

Here $D = V/8\pi^3$ and $V$ is the volume.

Notice that the energy has a minimum for $\Delta > 0$ only if $\int_0^\pi \Phi(\vec{K}_F, \vec{K}_F')\sin(\theta)d\theta < 0$

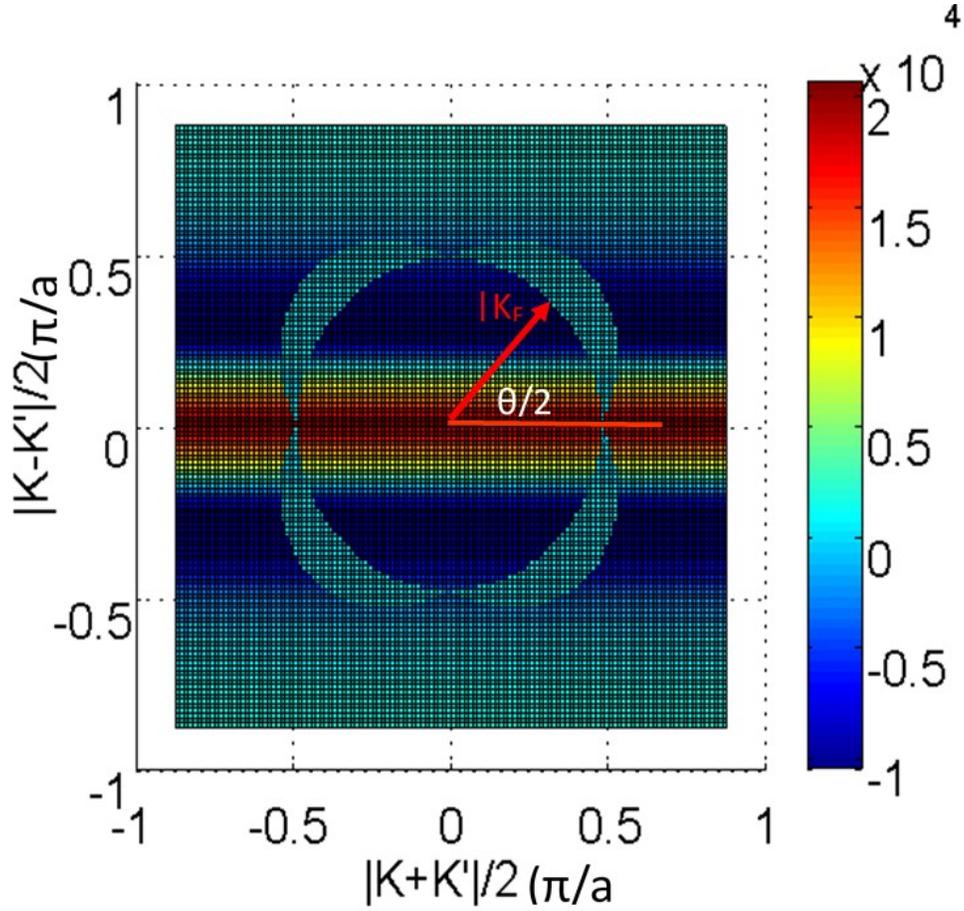

Figure 6. $\Phi$ as a function of $|\vec{K} + \vec{K}'|/2$ and $|\vec{K} - \vec{K}'|/2$ for $|K| = |K'|$; light blue shape $\sin(\theta)$.

In Fig. 6 we show $\Phi$ as a function of $|\vec{K}_F + \vec{K}_F'|/2$ and $|\vec{K}_F - \vec{K}_F'|/2$ for $|K_F| = |K_F'|$. Notice that unlike the case of classical superconductors $\Phi$ has both positive and negative values. We also show in Fig. 5 $\sin(\theta)$ and it can be seen that for the value of $K_F$ illustrated in the figure $\int_0^\pi \Phi(\vec{K}_F, \vec{K}_F')\sin(\theta)d\theta < 0$.

We have calculated the gap energy $\Delta_{K_F}$ as a function of the doping level from 0.084 to 0.4. In Fig. 1 we present the screening factors. In this example we assume that the polarizability is large in the plane normal to the line joining the B perovskite atoms and is equal to $\chi = 5.6 \times 10^{-40} \, Cm^2/V = 5.040 \, \text{Å}^3$. This choice yields reasonable values for the gap energy. We further assume that the first carrier is located on an oxygen ion and the potential seen by the second carrier as a function of position along the arrow shown in Fig. 1 is shown in Fig. 7 for 3 carrier densities.

The gap energy obtained under these conditions is shown in Fig. 8. Notice that the gap is zero at very low and at very high doping levels. The gap rises abruptly as the doping increases and diminishes gradually as the doping increases further.

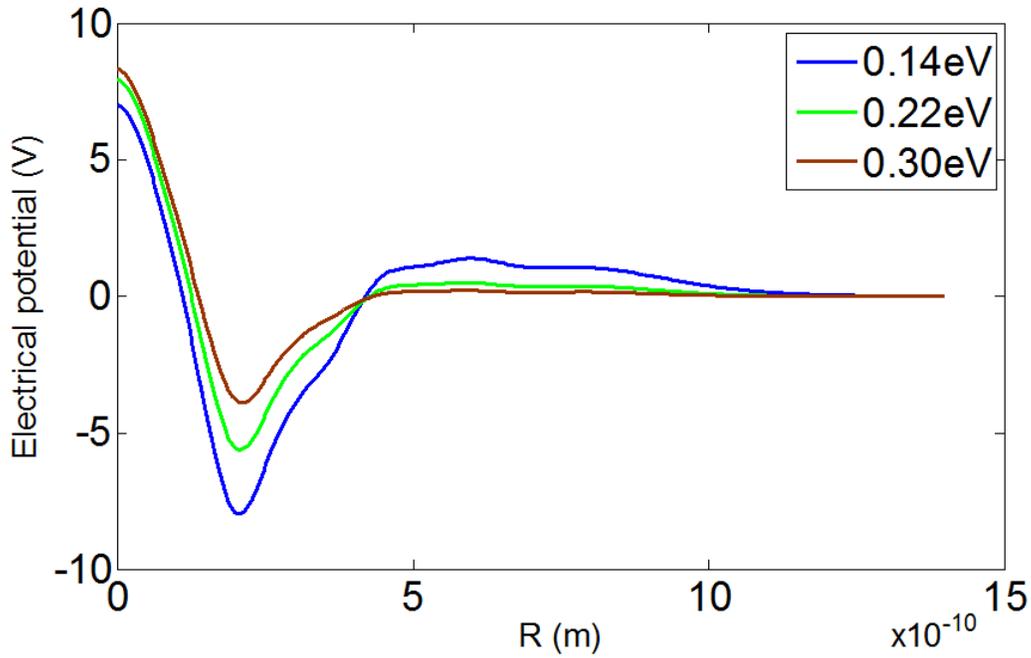

Figure 7. Electrical potential seen by second carrier as a function of distance along the arrow shown in Fig. 2 for 3 Fermi energies 0.14, 0.22, 0.30 eV that correspond to 0.1, 0.21, 0.33 carriers per unit cell doping levels.

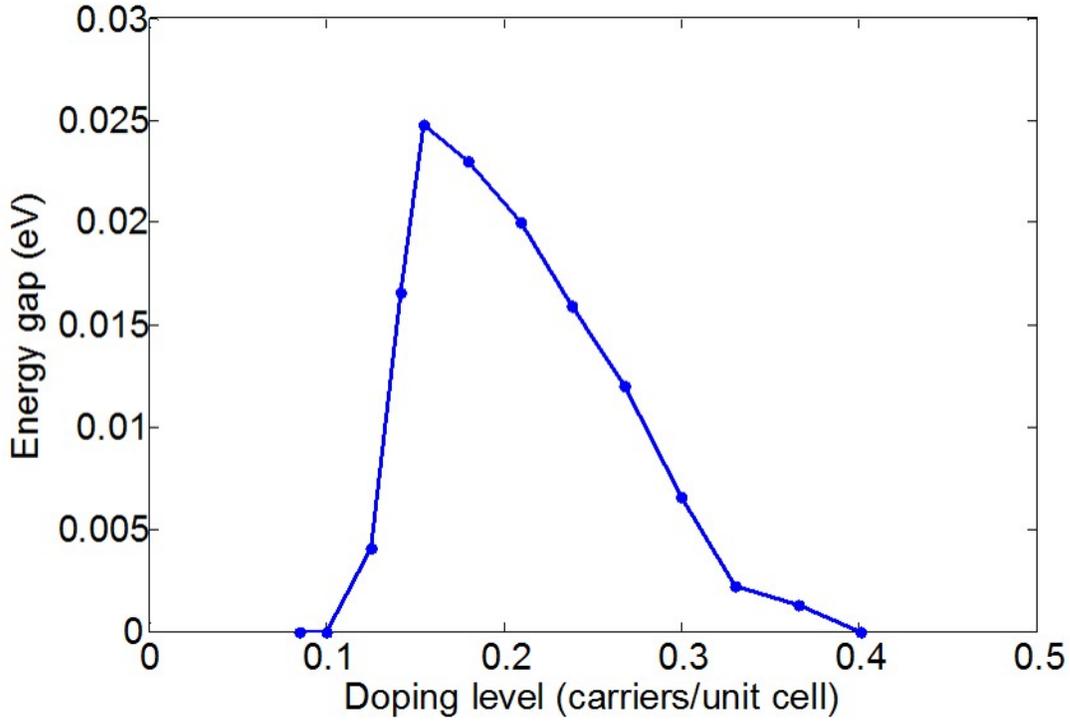

Figure 8. Energy gap as a function of doping level.

## 4. Discussion

The model we have presented here suggests that ion electronic polarizability can serve as a pairing mechanism in high Tc superconductors. Accurate calculation of the properties of these superconductors based on this mechanism is complicated and requires very heavy computations. Therefore we shall discuss these properties here only qualitatively and detailed quantitative calculations will be carried out later on.

As mentioned in sections 2 and 3 the polarizabilities must be large enough to produce an attractive potential between the charge carriers. As shown in Fig. 4 polarizabilities in the rage of 5 to 6.3 $Å^3$ that are consistent with measured polarizabilities in oxides and pnictides [16,17] yield pairing energies relevant to high Tc superconductors.

In high Tc superconductors the superconducting phase is limited on both low and high doping levels. The fact that the superconducting temperature decreases at higher doping is surprising because in classical superconductors Tc increases with increasing density of states. This can be understood in our model. At low carrier densities $K_F$ is small. From Fig. 6 it can be easily seen that $\int_0^\pi \Phi(\vec{K}_F, \vec{K}_F')\sin(\theta)d\theta > 0$ so $\Delta_{K_F} = 0$. Physically this means that wavefunctions with long wavelengths cannot confine the carriers to the regions of negative potential in spite of the fact that the potential well is deep. As soon as the wavelength becomes short enough the energy gap jumps to a large value because the screening length is large giving rise to large electric fields at the neighboring oxygen ions creating a deep potential well. Then as the doping increases the screening length becomes shorter and the potential well decreases decreasing the gap energy. This behavior is clearly seen in Fig. 8 and it closely resembles the experimentally observed behavior in cuprates[9] (panel 4). A similar behavior is seen in the bismutates as a function of K doping[24]. The onset of superconductivity may be a result of the transition to the cubic phase but the rapid decrease in Tc seems to be a result of the increasing screening which decreases the polarization of the oxygen ions.

The isotope effect in high Tc superconductors is generally small compared to the isotope effect in classical superconductors ($\alpha = 0.5$) [25,26] and in some instances it is even negative[27]. The change in mass of the ions affects the zero-point vibration amplitude. In optical modes the ions vibrate against each other with an amplitude comparable to the displacement of the electronic cloud under an electric field due to a charge carrier. So this will probably have a very small but finite effect on the polarizability. However a small change in the polarizability has a rather large effect on the energy gap $\Delta$. For example, changing the polarizability from $\chi = 5.04$ to $5.4 \text{Å}^3$ changes $\Delta$ from 0.0248 to 0.04eV. So in principle it seems that this mechanism may contribute to the observed isotope effect.

The strong dependence of the gap on polarizability raises the question why all the known high Tc superconductors and gap energies are in a rather narrow range of about 150K. The reason seems to be as follows: if the polarizability increasing beyond a certain value the system undergoes a dielectric catastrophe namely the electric fields are no longer linear with the carrier charge. As seen in Fig. 4 this happens in the present model when $\chi > 6.3$ $\text{Å}^3$ and pairing energies comparable to those observed in high Tc's can be reached with smaller but not much smaller polarizabilities.

The distance between paired carriers in classical superconductors is of the order of 100Å. In high Tc superconductors it is of the order of 10Å[9]. This is consistent with the distance suggested by our model (see Fig. 4). The fact that the pair distance is very small and depends only on very short distance interactions may also help explain very interesting recent observations of high Tc superconductivity in ultrathin films[28].

## 5. Summary and conclusions.

Based on simplistic calculations we have shown that rather large electronic polarizabilities of the oxygen ions in the cuprates and bismutates and the As or Se ions in the iron pnictides may serve as the pairing mechanism in these superconductors. We have shown that Cooper like pairs may form with energies smaller than the Fermi energy. In the case discussed here the energy difference is 0.011eV=132K and it can be changed up or down by changing the polarizability. The calculated pair wave function shows that under the conditions assumed here the distance between the carriers is about 1nm. Assuming that a BCS like wave-function describes the multi carrier system we have calculated the energy gap as a function of doping and showed that for $O^{2-}$ experimentally measured polarizabilities the energy gap dependence on doping closely resembles the measured gap. This similarity is inherent to the polarizability model.

The calculations presented here are not accurate enough to be conclusive. The calculations need to be improved in a number of ways:

1. The two carrier interaction matrix $\Phi$ needs to be calculated using the one carrier Bloch functions rather than just the plane waves we have used here.
2. $\Phi$ is not isotropic and the anisotropy needs to be taken into account.
3. We assume that the polarizability is constant independent of the electric field. The field dependence of the polarizability may be important.
4. In principle the polarizability is not a scalar. It may be different along the line connecting two cations and the directions perpendicular to it.
5. Quadrupole polarizability may also play a role.

Taking all these points into account will require rather lengthy calculations that will be done in the future.

**Appendix 1**

The repulsive electric field is given by

$$E_R = \frac{e + e_s}{4\pi\varepsilon_0 |s|^2} \tag{A1}$$

where $\vec{s} = \vec{r}_2 - \vec{r}_1$ and $e_s$ is the screening charge within a sphere of radius $s$ around the charge.

$$e_s = 4\pi \int_0^s \rho_s(s') s'^2 \, ds' \tag{A2}$$

Where the charge density

$$\rho_s(s) = \int eD(\varepsilon)\{[1+\exp[(\varepsilon - \varepsilon_F + e\varphi_R)/k_B T]^{-1} - [1+\exp[(\varepsilon - \varepsilon_F)/k_B T]^{-1}\}d\varepsilon \tag{A3}$$

$D(\varepsilon)$ is the density of states, $\varepsilon$ is energy measured relative to the band extremum and

$$\varphi_R(s) = \int_s^\infty \vec{E}_R(s') d\vec{s}' \tag{A4}$$

Close to the charge carrier the potential is very large (a number of eV) so we can approximate the charge density by the acceptor (donor) density $\rho_0$

The polarizability induced potential can be estimated by

$$\vec{E}_p(\vec{r}_1,\vec{s}) = \frac{1}{4\pi\varepsilon_0} \sum_{j\neq 1}[\frac{\vec{p}_j}{|\vec{s}_j-\vec{s}|^3} - \frac{3(\vec{s}_j-\vec{s})[\vec{p}_j \bullet (\vec{s}_j-\vec{s})]}{|\vec{s}_j-\vec{s}|^5}] \qquad A(5)$$

We have neglected here the effect of screening because the electric field decreases with the distance to the third power and the relevant region is practically depleted of free carriers. The polarization needs to be calculated self-consistently:

$$\vec{P}_j = \chi\vec{E}(\vec{r}_1,\vec{s}_j) = \chi[\vec{E}_R(\vec{s}_j) + \vec{E}_P(\vec{r}_1,s_j)] \qquad A(6)$$

and $\chi$ is the polarizability tensor.

The electric field at point $\vec{s}$

$$\vec{E}(\vec{r}_1,\vec{s}) = \vec{E}_R(\vec{s}) + \vec{E}_P(\vec{r}_1,s)$$

and the electrical potential induced by the charge carrier is

$$\hat{\varphi}(\vec{r}_1,\vec{s}) = \int_{\vec{s}}^{\infty} \vec{E}(\vec{r}_1,\vec{s}') \bullet d\vec{s}'.$$

In terms of the pair center of mass $\vec{R} = (\vec{r}_2+\vec{r}_1)/2$ and inter-pair distance $\vec{S} = (\vec{r}_2-\vec{r}_1)/2$

$$\varphi(\vec{R},\vec{S}) = \hat{\varphi}(\vec{r}_1,\vec{s}) \qquad A(7)$$

## Acknowledgments

The authors gratefully acknowledge useful discussions with Dror Orgad from the Hebrew University of Jerusalem.